# A comparative study on instability of steady flows in helical pipes

*A. Yu. Gelfgat*

*School of Mechanical Engineering, Faculty of Engineering, Tel-Aviv University, Ramat Aviv, Tel-Aviv 69978, Israel. e-mail: gelfgat@tau.ac.il*

**Abstract**

A computational study of three-dimensional instability of steady flows in a helical pipe of arbitrary curvature and torsion is carried out for the first time. The problem is formulated in Germano coordinates in two equivalent but different forms of the momentum equation, so that results obtained using both formulations cross verify each other. An additional formulation in the cylindrical coordinates is applied for a limiting case of the toroidal pipe. The calculations are performed by the finite volume and finite difference methods. Grid independence of the results is established for both steady flows, the eigenvalues associated with the linear stability problem, and the critical parameters. The calculated steady flows agree well with experimental measurements and previous numerical results. The computed critical Reynolds numbers corresponding to the onset of oscillatory instability agree well with the most recent experimental results, but disagree with the earlier ones. Novel results related to the parametric stability study are reported.

## 1. Introduction

A helical pipe is widely used as an effective mixing tool (Vadhisth et al 2008). The Dean vortices necessarily appear in the helical pipe flow driven by an applied pressure drop regardless the flow intensity. These vortices effectively mix either heat or mass without any need of additional mixing means. Contrarily to flows in straight ducts, circular or rectangular, no analytic solutions, similar to the Poiseuille profile, can be found for the helical pipe flow. Therefore, the numerical modeling is called for, even at low and moderate values of the Reynolds number. In this manuscript we focus mainly on examination of stability of calculated steady flows and computation of critical parameters at which the primary transition takes place.

For a long time, the helical pipe flow was considered as three-dimensional until Germano (1982) showed that a two-dimensional formulation is possible in a specially tailored system of curvilinear orthogonal coordinates. Since then, computations of this flow became affordable (see, e.g., Yamamoto et al 1994; Hüttl & Friedrich, 2000; Gelfgat et al 2003; Nobari & Malvandi, 2013; Totirean et al. 2016; and references therein). However, to the best of the author's knowledge, no comparison between independent numerical results was ever reported. It seems also that no quantitative comparison between experimentally observed and computed steady flows was ever published. Such comparisons are mainly qualitative and compare experimentally visualized flow patterns with graphical representations of calculated flows (see, e.g., Yamamoto et al, 2002). A not very successful comparison between measured and calculated turbulent flow was reported by Webster & Humphrey (1997). Apparently, an accurate calculation of steady flow states is a necessary precondition for further stability studies. In the following, we address also this issue.

Experimental studies of onset of instability in a helical pipe flow start from works of White (1929) and Taylor (1929). Later measurements of the instability threshold were carried out by Sreenivasan & Strykowski (1983), and Webster & Humphrey (1993). All these papers considered pipes with a rather small torsion. The measured critical Reynolds numbers reported in these works exhibited a considerable scatter for similar pipe curvatures. In recent experiments of Kühnen et al (2014, 2015), the dimensionless curvature was varied between 0.03 and 0.08, and the torsion also was very small. The critical Reynolds numbers measured in this study are significantly smaller than those reported in the previous works, and are partially supported by the numerical results of Canton et al (2016) obtained for a toroidal pipe. The effect of a large torsion on flow stability was studied by Yamamoto et al (1995), who reported destabilization of flow with increase of the torsion from the zero value, followed by a slight stabilization at larger torsions. Later Yamamoto et al (1998) studied the instability onset numerically for parameters of the latter experiment, but considering only two-dimensional perturbations. Only qualitative agreement between experimental and numerical critical Reynolds numbers was established.

Most of the numerical studies of stability of steady flows in helical pipes addressed either only two-dimensional disturbances (Yamamoto et al., 1998), or only toroidal geometry (Webster & Humphrey, 1997; Di Piazza & Ciofalo, 2011; Canton et al, 2016). These approaches are apparently incomplete. It should be noted that the toroidal pipe geometry is a mathematical limit corresponding to the helical pipe with a zero torsion. For a long time it was considered to be a purely mathematical model, but Kühnen et al (2014) showed that such geometry can be realized in experiment.

Fully three-dimensional time-dependent computations in a helical pipe (see, e.g., Hüttl & Friedrich, 2000) are quite rare and are better fitted for modelling of strongly supercritical flows with large oscillation amplitudes. These approaches may fail near stability limits. To calculate critical parameters, at which steady flow becomes unstable, mathematically rigorous approaches based on the analysis of the spectrum of the linearized equations are more pertinent. Thus, Canton et al (2016) studied linear stability of a flow in a toroidal pipe with respect to all possible infinitesimal three-dimensional perturbations and reported critical Reynolds numbers close to those reported in the experimental studies of Kühnen et al (2014, 2015).

In this paper we focus on an accurate computation of steady flows in helical pipes and computing of critical parameters corresponding to the onset of their primary instability. The results are obtained using two different formulations of governing equations in Germano coordinates. For calculation of steady states and the linear stability studies, we apply the methodology of Gelfgat (2007). The results obtained for each formulation are convergent and agree with each other. To validate computation of steady flows at moderate Reynolds numbers against independent computational and experimental data, we start from comparison with the numerical results of Yamamoto et al. (1994). We report a successful comparison and add some more fully converged data that can be used for benchmarking. To establish agreement with the experimental measurements, we compare measured and calculated friction factors, and show that at moderate Reynolds numbers, the present result agree well with the experimental measurements of De Amicis (2014).

After the correctness of the calculated steady flows is established, we consider their linear stability with respect to all possible three-dimensional infinitesimal perturbations. The main difficulty in this task is calculation of leading eigenvalues of the linearized stability problem. We use parameters of Yamamoto et al. (1994) to study the convergence of the eigenvalues, and then the convergence of the critical parameters of the problem. The successful convergence and comparison exercises allow to confidently report the first numerical results on three-dimensional linear stability of steady flows in helical pipes with arbitrary curvature and torsion.

In the following, we successfully compare our stability results with the recent experiments of Kühnen et al (2014, 2015), as well as with the recent numerical results for the toroidal pipe of Canton et al (2016). However, our attempts to compare with the earlier experimental stability studies of White (1929), Taylor (1929), Sreenivasan & Strykowski (1983), and Webster & Humphrey (1993), led to noticeably smaller computed critical Reynolds numbers than those reported in the experiments. All the above experiments and computations were carried out for pipes with small torsion. The critical Reynolds numbers calculated for the curvatures and torsions of these experimental setups appeared to be very close to the results of Kühnen et al (2014, 2015), which, again, are noticeably smaller than those reported previously. We also compared our numerical results with the experimental results of Yamamoto (2005) obtained for helical pipes with large torsions. The critical Reynolds numbers computed for the curvatures and torsions corresponding to these experiments are either close to the experimental ones, or are below them.

Finally, we report a stability diagram calculated for a helical pipe with fixed curvature and varying torsion. We show that with the increase of torsion, the instability sets in due to different most unstable eigenmodes that represent distribution of amplitudes of oscillations developing in a supercritical regime. A complete parametric stability study for all relevant values of the curvature and the torsion would require not only many computational runs, but also abundant graphical material. This is beyond the scope of present study.

We conclude this paper by summarizing successful comparison exercises and some novel results, and discuss possible sources of disagreement between the measured and calculated critical parameters.

## 2. Coordinate system

To describe the system of orthogonal curvilinear coordinates introduced by Germano (1982), we start from definitions related to the pipe centerline, which is a helical curve defined parametrically as

$$\boldsymbol{R}_0(t) = \{x(t), y(t), z(t)\} = \{c \cdot cos(t), c \cdot sin(t), bt\}, \qquad (1)$$

where $c$ is the radius of the helix, and $2\pi b$ is distance between coils (see Fig. 1). The curvature and the torsion of the helical curve are defined as

$$\kappa = \frac{c}{b^2+c^2}, \quad \tau = \frac{b}{b^2+c^2}, \qquad (2)$$

respectively. In the following we use also their ratio $\lambda = \tau/\kappa = b/c$.

The most natural way to define the position of the point inside a helical pipe with circular cross-section was proposed by Wang (1981). The point position is defined by its location $s$ with respect to the pipe centerline and by polar coordinates $(r, \theta)$ inside a pipe cross-section orthogonal to the centerline (Fig. 1). However, as shown by Wang (1981) and Germano (1982), these coordinates appear to be non-orthogonal. This inconvenience was removed by Germano (1982), who proposed to rotate the position of $\theta = 0$ along with the pipe centerline as (assuming $\tau$ is a constant)

$$\xi = \theta - \int_{s_0}^{s} \tau ds = \theta - \tau(s - s_0) \tag{3}$$

The resulting coordinate system $(r, \xi, s)$ is orthogonal. The Lamé coefficients of these coordinates are $H_r = 1$, $H_\xi = r$, $H_s = 1 + kr\sin(\xi)$. Note that for the constant torsion

$$\frac{d}{ds} = \frac{\partial}{\partial s} + \frac{\partial}{\partial \xi}\frac{\partial \xi}{\partial s} = \frac{\partial}{\partial s} - \tau\frac{\partial}{\partial \xi} = \frac{\partial}{\partial s} - \kappa\lambda\frac{\partial}{\partial \xi} \tag{4}$$

In the coordinates $(r, \xi, s)$ we can assume that the three fluid velocity components and the pressure are independent on the position at the pipe centerline, i.e., independent on $s$, so that $\partial/\partial s = 0$. Thus, we arrive to a two-dimensional formulation for the velocity and the pressure dependent only on $r$ and $\xi$.

## 3. Governing equations

### 3.1. General case in Germano coordinates

We consider a flow of incompressible fluid in a helical pipe of the inner radius $a$, radius of the coil $c$, and a constant distance between the coils equal to $2\pi b$. The pipe is sketched in Fig. 1. The flow is created by a pressure gradient, which is constant along the pipe centreline

$$\frac{dP}{ds} = G = const, \tag{5}$$

and is governed by the continuity and momentum equations. The flow is characterized by the three dimensionless parameters, which are dimensionless curvature $\varepsilon = a\kappa$, torsion to curvature ratio $\lambda$, and the Reynolds number $Re_d = 2a\,\overline{U}/\nu$, where $d = 2a$ is the pipe diameter, $\nu$ is the kinematic viscosity, and $\overline{U}$ is the flow mean velocity. The Reynolds number sometimes is replaced by the Dean number $De_d = Re_d\sqrt{\varepsilon}$.

The above definition of the Reynolds (Dean) number requires mean velocity value, which is convenient for experimental studies. In a numerical study, the mean velocity can be found only after calculation of the flow. Since it is not known a priory, its use in the problem formulation causes certain inconvenience. Thus, to use this traditional formulation, a non-linear problem making dimensionless $\bar{U}$ equal to unity was solved in Canton et al. (2016). To make an alternative and more convenient non-dimensionalization, we use the pressure gradient based scales introduced in Gelfgat et al. (2003). Assuming that the pressure gradient $G$ is known, we define the scales of length, time, velocity and pressure as $a, (\rho a/G)^{1/2}, (Ga/\rho)^{1/2}$, and $Ga$. The resulting system of the dimensionless continuity and momentum equations reads

$$\nabla \cdot \mathbf{v} = 0, \quad \frac{\partial \mathbf{v}}{\partial t} + (\mathbf{v} \cdot \nabla)\mathbf{v} = -\frac{G}{H_s}\mathbf{e}_s - \nabla p + \frac{1}{R_G}\Delta \mathbf{v} \qquad (6,7)$$

where the dimensionless parameter $R_G = (Ga^3/\rho\nu^2)^{1/2}$ replaces the Reynolds number. The equations (6,7) are solved together with the no-slip condition

$$\mathbf{v}(r = a, \xi, s) = 0. \qquad (8)$$

After the flow is computed, its dimensionless mean velocity $\bar{V}_G$ can be easily obtained. Then the dimensional mean velocity is $\bar{U} = \bar{V}_G(Ga/\rho)^{1/2}$, and the resulting Reynolds number is calculated as $Re_d = 2a\,\bar{U}/\nu = 2a\,\bar{V}_G(Ga/\rho)^{1/2}/\nu = 2\bar{V}_G R_G$.

The friction factor is defined as ($L$ is the pipe length)

$$f = \frac{\Delta p}{\rho \bar{U}^2/2}\left(\frac{2a}{L}\right) = 4\frac{\Delta p}{L}\frac{a}{\rho \bar{V}_G^2(Ga/\rho)} = \frac{4}{\bar{V}_G^2} \qquad (9)$$

For a "two-dimensional" flow independent on the coordinate $s$, the continuity equation in the helical coordinates reads

$$\nabla \cdot \mathbf{v} = \frac{1}{rH_s}\left\{\frac{\partial}{\partial r}[rH_s v_r] + \frac{\partial}{\partial \xi}[H_s v_\xi] - \varepsilon\lambda r\frac{\partial v_s}{\partial \xi}\right\} = 0. \qquad (10)$$

It can be satisfied by introducing a function $\hat{\psi}$ as

$$rH_s v_r = \frac{\partial \hat{\psi}}{\partial \xi}, \quad H_s v_\xi - \varepsilon\lambda r v_s = -\frac{\partial \hat{\psi}}{\partial r}, \qquad (11)$$

This is not a "real" stream function since $\mathbf{v} \neq rot[\hat{\psi}\mathbf{e}_s]$. As is noted by several authors (e.g., Germano, 1982; Yamamoto et al 1994), the quasi-two-dimensional flows are defined by two scalar functions $v_s(r,\xi)$ and $\hat{\psi}(r,\xi)$. To make the function $\hat{\psi}$ more meaningful physically, we modify it as $\psi = \hat{\psi}/H_s$, which leads to

$$v_r = \frac{1}{rH_s}\frac{\partial[H_s\psi]}{\partial \xi}, \quad v_\xi - \frac{\varepsilon\lambda r}{H_s}v_s = -\frac{\partial[H_s\psi]}{\partial r}, \qquad (12)$$

so that now

$$v_r \mathbf{e}_r + \left[v_\xi - \frac{\varepsilon\lambda r}{H_s}v_s\right]\mathbf{e}_\xi = rot[\psi\mathbf{e}_s], \qquad (13)$$

and function $\psi$ can be interpreted as the stream function of the field $\left(v_r, v_\xi - \frac{\varepsilon\lambda r}{H_s}v_s, 0\right)$. In the following, we call $\psi$ pseudo – streamfunction.

The momentum equations are written in general orthogonal coordinates as in Kochin et al. (1954)

$$\frac{\partial v_1}{\partial t} + \frac{v_1}{H_1}\frac{\partial v_1}{\partial x_1} + \frac{v_2}{H_2}\frac{\partial v_2}{\partial x_2} + \frac{v_3}{H_3}\frac{\partial v_3}{\partial x_3} + \frac{v_1 v_2}{H_1 H_2}\frac{\partial H_1}{\partial x_2} + \frac{v_1 v_3}{H_1 H_3}\frac{\partial H_1}{\partial x_3} - \frac{v_2^2}{H_1 H_2}\frac{\partial H_2}{\partial x_1} - \frac{v_3^2}{H_1 H_3}\frac{\partial H_3}{\partial x_1} = -\frac{1}{H_1}\frac{\partial p}{\partial x_1} +$$

$$+ \frac{1}{R_G}\frac{1}{H_2 H_3}\left\{-\frac{\partial}{\partial x_2}\left[\frac{H_3}{H_1 H_2}\frac{\partial (H_2 v_2)}{\partial x_1}\right] + \frac{\partial}{\partial x_2}\left[\frac{H_3}{H_1 H_2}\frac{\partial (H_1 v_1)}{\partial x_2}\right] + \frac{\partial}{\partial x_3}\left[\frac{H_2}{H_1 H_3}\frac{\partial (H_1 v_1)}{\partial x_3}\right] - \frac{\partial}{\partial x_3}\left[\frac{H_2}{H_1 H_3}\frac{\partial (H_3 v_3)}{\partial x_1}\right]\right\}. \quad (14)$$

Here the indices 1, 2 and 3 stay for $r$, $\xi$, and $s$, respectively. Two other equations are obtained by cyclic permutations of the indices. These equations contain mixed second derivatives, which may cause certain inconvenience and loss of accuracy at the discretization stage. To avoid this, the mixed derivatives are eliminated using the continuity equation (10). The resulting set of momentum equations is detailed in the Appendix. In the following, applying eqs. (14) is called Formulation 1, while applying eqs. (A1)-(A6) from the Appendix is called Formulation 2. Mathematical equivalence of both formulations is verified by symbolic computations on a computer.

### 3.2. Flow in a torus: a particular case in cylindrical coordinates

To have a fully independent calculation for comparison purposes, we consider also a problem of flow in a toroidal pipe driven by a constant azimuthal pressure gradient $G_\varphi = \partial p/\partial \varphi$ (see also Gelfgat et al, 2003; Canton et al, 2016). The problem geometry is sketched in Fig. 2. The torus center is placed in the origin of the cylindrical coordinates $(\hat{r}, \varphi, z)$ so that its center line lies in the plane $z = 0$. The center of the torus cross-section is located at the distance $R$ from the origin. Radius of the torus cross-section is $a$. The border of the torus is defined by the line $(x - R)^2 + z^2 = a^2$. For dimensionless formulation similar to the helical pipe, scales of the length, time, velocity and pressure are $a, a(\rho/G_\varphi)^{1/2}, (G_\varphi/\rho)^{1/2}$, and $G_\varphi$. Thus, for the dimensionless formulations, the parameters of Fig. 2 must be replaced by $a = 1$, $R = 1/\varepsilon$, where $\varepsilon = a/R$ is the dimensionless curvature of the torus axis. The dimensionless momentum equation reads

$$\frac{\partial \boldsymbol{v}}{\partial t} + (\boldsymbol{v} \cdot \nabla)\boldsymbol{v} = -\frac{1}{r}\boldsymbol{e}_\theta - \nabla p + \frac{1}{R_\varphi}\Delta \boldsymbol{v} \, , \quad (15)$$

where $R_\varphi = \left(G_\varphi a^2/\rho \nu^2\right)^{1/2}$. Comparing with the formulation in the Germano coordinates we notice that

$$G = \frac{\partial p}{\partial s} = \frac{1}{R}\frac{\partial p}{\partial \theta} = \frac{1}{R}G_\varphi \, . \quad (16)$$

Thus,

$$R_\varphi = \sqrt{\frac{G_\varphi \rho a^2}{\mu^2}} = \sqrt{\frac{GR\rho a^3}{\mu^2 a}} = \frac{Re_G}{\sqrt{\varepsilon}} \quad . \tag{17}$$

Obviously, the velocity and time scales are also related by $\sqrt{\varepsilon}$. As above, the average velocity based Reynolds number is $Re_d = 2a\bar{U}/\nu = 2a\bar{V}_{G_\varphi}(G_\varphi/\rho)^{1/2}/\nu = 2\bar{V}_{G_\varphi}R_\varphi$. The transformations of velocity compomemts from cylindrical to Germano coordinates and back are defined by

$$v_s = \frac{1}{\varepsilon \hat{r}}(1 + \varepsilon r \sin\varphi)v_\varphi, \tag{18.1}$$

$$v_r = \frac{1}{r}\left[(\hat{r} - R)\left(\dot{d}\cos\varphi - \frac{d}{\hat{r}}v_\theta \sin\varphi\right) + zv_z\right] \quad , \tag{18.2}$$

$$v_{\xi_s} = \frac{1}{r}\left[(\hat{r} - R)v_z - z\left(\dot{d}\cos\varphi - \frac{d}{\hat{r}}v_\varphi \sin\varphi\right)\right] \quad , \tag{18.3}$$

$$v_{\hat{r}} = v_r \cos\xi - v_\xi \sin\xi, \quad v_\theta = v_r \sin\xi + v_\xi \cos\xi, \quad v_z = \frac{\varepsilon(R + r\cos\xi)}{1 + \varepsilon r \sin\theta} v_s \,, \tag{19}$$

where $= \sqrt{\hat{r}^2 + z^2}$, $\dot{d} = (\hat{r}v_{\hat{r}} + zv_z)/d$.

## 4. Numerical technique

The general problem formulated in the Germano coordinates was solved on staggered grids using central finite differences with linear interpolation between the nodes where necessary. The Newton iteration was applied for calculation of steady flows. Application of the Newton method is identical to Gelfgat (2007), and is based on the LU decomposition of the sparse Jacobian matrix with further analytic solution for the Newton corrections.

Consideration of the linear stability of calculated steady flows requires taking into account infinitesimally small disturbances that can be periodic along the pipe centerline direction $s$. The perturbations were represented in the form $\{\tilde{v}(r,\xi), \tilde{p}(r,\xi)\}exp[\sigma t + iks]$, where $\sigma$ is the complex time increment, $k$ is the wavenumber along the centerline and infinitesimally small perturbation amplitude is denoted by tilde. The linearization procedure is standard, except of derivatives in the $s-$ direction, for which eq. (4) must be replaced, for the dimensionless variables, by

$$\frac{d}{ds} = -\varepsilon\lambda\frac{\partial}{\partial \xi} + ik \quad . \tag{20}$$

The linear stability problem reduces to the generalized eigenvalue problem

$$\sigma \mathbf{B}(\tilde{v}, \tilde{p})^T = \mathbf{J}(\tilde{v}, \tilde{p})^T \tag{21}$$

Here **J** is the Jacobian matrix that defines r.h.s. of the linearized problem and **B** is the diagonal matrix such that its diagonal elements corresponding to the time derivatives of $\tilde{v}$ are equal to one, while the elements corresponding to $\tilde{p}$ and the boundary conditions are zeros, so that $det\mathbf{B} = 0$. Thus, the generalized eigenproblem (21) cannot be transformed into a standard. To study stability of an axisymmetric steady flow state for a given set of the governing parameters, it is necessary to compute the eigenvalue $\hat{\sigma}$ having the largest real part for all real wavenumbers $k$. This $\hat{\sigma}$ is called leading eigenvalue. Apparently $\hat{\sigma} = \max_{k}\{Real[\sigma(k)]\} > 0$ means instability of the axisymmetric steady flow state. The value of the wavenumber yielding the maximum of $Real[\sigma(k)]$ is called critical and is denoted as $k_{cr}$. The imaginary part of the leading eigenvalue is called critical frequency and is denoted as $\omega_{cr} = Im[(k_{cr})]$. The corresponding eigenvector of (21) is called leading. It defines the most unstable perturbation of the base state. Its amplitude, to within multiplication by a constant, represents the amplitude of an oscillatory flow resulting from the instability onset.

The eigenproblem (21) is solved by the Arnoldi iteration in the shift-and-invert mode

$$(\mathbf{J} - \sigma_0 \mathbf{B})^{-1} \mathbf{B}(\tilde{v}, \tilde{p})^T = \vartheta(\tilde{v}, \tilde{p})^T, \quad \vartheta = \frac{1}{\sigma - \sigma_0} \quad (22)$$

where $\sigma_0$ is a complex shift. The Arnoldi method realized in the ARPACK package of Lechouq et al. (1998) is used. Following Gelfgat (2007) we calculate LU decomposition of the complex matrix $(\mathbf{J} - \sigma_0 \mathbf{B})^{-1}$, so that calculation of the next Krylov vector for the Arnoldi method is reduced to one backward and one forward substitutions. It should be noted that the Jacobian matrices for the Newton iteration and the stability analysis are different, since it contains the terms depending on the wavenumber $k$ that can also be complex. The Jacobian matrices were calculated directly form the numerical schemes. The corresponding parts of the code were verified by numerical differentiation of the equations' right hand sides.

To calculate the leading eigenvalue $\hat{\sigma}$ it is necessary to choose the shift $\sigma_0$ close to $\hat{\sigma}$ and to calculate 10-20 eigenvalues with the largest absolute value. In the following calculations, we fix $Real(\sigma_0)=0$ and vary $Im(\sigma_0)$ until the leading eigenvalue $\hat{\sigma}$ is computed. Then we calculate the instability point with $\hat{\sigma} = (0, Im(\hat{\sigma}))$ and vary $Im(\sigma_0)$ further to ensure that there is no other eigenvalue with a larger real part. After that we vary the wavenumber $k$ to find at which $k = k_{cr}$ the instability takes places at the lowest $R_G = R_{G,cr}$. At this stage we apply the golden ratio algorithm. For a given pair of the geometrical governing parameters $\varepsilon$ and $\lambda$, the result of the stability study is defined by the critical values $R_{G,cr}$, $k_{cr}$, $\omega_{cr}$ and the leading eigenvector. The critical Reynolds number then can be calculated as $Re_{d,cr} = 2\bar{V}_G R_{G,cr}$, and the dimensionless critical frequency scaled by $(2a)/\bar{U}$ is $\omega_{d,cr} = 4\omega_{cr} R_{G,cr}/Re_{d,cr}$.

To gain an additional validation of the results, we consider a limiting case of zero torsion, which brings us to flow in a toroidal pipe considered by Catton et al (2016). This case is considered in general Germano coordinates, as well as in the cylindrical coordinates, as described above. This allows us to compare our results obtained by the two independent approaches and compare them with the independent calculations of Canton et al (2016). Unfortunately, it can be done only for the zero torsion. For calculation in the cylindrical coordinates, the momentum and continuity equations are discretized by the finite volume method. The toroidal boundary is treated by the immersed boundary method using the approach of Kim et al (2001). Then the steady states and their stability are treated as in Gelfgat (2007). In this case, the perturbation is assumed to be a circumferential Fourier mode $\sim exp(im\theta)$. Since $m$ is integer, values of the wavenumber $k$ are discrete and are defined by $k = m/\varepsilon$. The critical $m$ and $k$ are found by screening several discrete values until a global minimum of $R_{G,cr}(m)$ is obtained.

## 5. Results

### 5.1. Steady flows: comparison with experiment

To gain an initial validation of the formulation, the numerical approach and the code, we compare with the experimental results of De Amicis et al. (2014). The computed and measured friction factors as functions of the Reynolds numbers are sown in Fig. 10 for four different helical pipes used in the experiments. It is seen that at Reynolds numbers lower than ≈3000, the calculations and the measurements coincide. At larger Reynolds numbers the results differ, which happen due to the laminar – turbulent transition, similarly to the straight pipes. At the same time, at $Re_d \leq 3000$ the non-linear terms already become significantly large, which allows us to assume that our code reproduces laminar flows correctly.

### 5.2. Steady flows: convergence and comparison with an independent solution

For the convergence study and comparison with independent results we chose the work of Yamamoto et al (1994), where ratios of mass fluxes in helical and straight pipes were reported for various curvatures and torsions. It can be easily seen that the mean velocity of the Poiseuille flow in a straight pipe is $\bar{w}_{Poisuille} = R_G/8$, so that the flux ratio is equal to $\bar{U}/\bar{w}_{Poisuille}$. The velocity in Yamamoto et al (1994) was rendered dimensionless using the viscous scale $\nu/a$, which resulted in the Dean number $Dn = R_G^2\sqrt{2\varepsilon}$. The two geometric parameters were $\delta = \varepsilon$ and $\beta = \tau/\sqrt{2\varepsilon}$. The solution was computed on $N_r \times N_\xi = 35 \times 60$ collocation points.

Comparison with the results of Yamamoto et al (1994) is shown in Table 1. We report results obtained using both formulations and uniform grids having $50 \times 100$ and $100 \times 200$ nodes. The Richardson extrapolations based on these grids and made for both formulations coincide to within the third decimal place at least. Taking into account that the grid used in Yamamoto et al. (1994) was rather coarse, the agreement with this work is considered as good. To allow for a more detailed comparison, the maximal and minimal values of the pseudo – streamfunction and the friction factor also are reported in Table 2, where we also observe an agreement to within the third decimal place for almost all reported values.

Additionally, we calculate the flow at $\delta = 0$ in a torus of the same curvature $\varepsilon = \beta = 0.4$. The results are reported in Table 3. These results verify those reported in Tables 1 and 2, but only for the zero torsion. The calculations in the cylindrical coordinates coincide with the previous ones to within the third decimal place starting from the $200^2$ uniform grid. At the same time, a convergence within four first decimal places requires very fine grids of $400^2$ and more nodes. Most probably, this slow convergence results from the immersed boundary approach which reduces the order of the whole numerical scheme to the first order.

The calculated steady flows are shown in Figs. 3 and 4. In the limit of zero torsion, the helical pipe turns into a torus, and the Dean vortices are symmetric. The third $s$-componen, of velocity, is advected by these vortices so that its maximum is shifted toward the outer side of the pipe. The Dean vortices symmetry is broken when the torsion is non-zero. With further growth of the torsion the two Dean vortices merge into a single one. Note that the meridional flow never vanishes since the non-zero pipe curvature always leads to a non-potential centrifugal force. An interesting observation is that with the increase of torsion, the meridional vortex monotonically intensifies, while the friction factor monotonically decreases (see Table 2).

*5.3.    Stability of steady flows: convergence studies*

Study of stability of steady flows is usually more computationally demanding than computation of steady flow states. This is also the case for flows in helical pipes, which is illustrated in the computations described below. We start from Table 4 that reports a convergence study carried out for the leading eigenvalues computed for the parameters of Yamamoto et al. (1994), and for the wavenumbers $k = 0$ and $0.8$. The calculations were carried out for the uniform grids of 50×100, 100×200, 200×400, and 300×600 nodes in the $r$ and $\xi$ directions, respectively. In the lower part of the table we report results obtained for the toroidal pipe ($\delta = \lambda = 0$) in the cylindrical coordinates for the azimuthal wavenumbers $m = 0$ and 2, which correspond to the above values of $k$.

At $Dn = 1000$, the Reynolds number $Re_d$ is relatively small and varies from 167.5 at $\delta = 0$ to 266.7 at $\delta = 2.5$. At such low Reynolds numbers the flow is expected to be stable. Actually, all the eigenvalues reported in Table 4 have negative real parts, meaning stability of the flows. At the same time, in the case $\delta = 1.5$ and $k = 0.8$, the real part of the leading eigenvalue already approaches zero. This indicates on a possible flow destabilization at large torsions, which is discussed below. Examination of the results reported in Table 4 shows that there are values of $\delta$ and $k$ at which we observe a good convergence and almost full coincidence between results obtained using both formulations. It happens, for example, at $\delta = 0$, at which the results of both formulations are almost identical and agree well with the results for the toroidal pipe. At the same time, there are cases, e.g., at $\delta = 1$, $k = 0$ and $> 1$, $k = 0.8$, for which the convergence is noticeably slower. Nevertheless, we observe two decimal places converged.

Convergence of all critical parameters, the critical number $R_G$, the critical Reynolds number $Re_d$, the critical oscillations frequency, and the critical wavenumber, is illustrated in Table 5. Again, we use the values of the curvature and torsion as in Yamamoto (1994). The lower part of the table shows convergence for the toroidal pipe, $\beta = \lambda = 0$, calculated in the cylindrical coordinates. At the zero torsion, the convergence of both formulations 1 and 2 is very good and the results coincide to within the fourth decimal place. Convergence of the results obtained in the cylindrical coordinates is noticeably slower, which possibly happens because of the immersed boundary description of the helical pipe wall. With the increase of torsion, the critical Reynolds numbers steeply decrease, so that at $\beta = 2.5$ it becomes approximately 50 times smaller than that at $\beta = 0$. Nevertheless, with the increase of $\beta$, the convergence slows down, so that for $\beta \geq 1$ only two decimal places are converged. This shows that at large torsions computational modelling becomes more demanding: the convergence should be monitored and finer grids are needed to reach a necessary accuracy.

*5.4. Stability of steady flows: comparison with experiment and independent computations*

Figure 6 shows comparison of the present stability results obtained for parameters of the recent experiments in the toroidal (Kühnen et al, 2014) and helical (Kühnen et al, 2015) pipes, and a recent stability study for toroidal pipes (Canton et al, 2016). Calculations for the toroidal pipe were performed using both formulations 1 and 2, and an additional formulation for torus in the cylindrical coordinates. Calculations were carried out on the uniform grid of 400×400 finite volumes in the cylindrical coordinates, and on the 400×800 uniform finite grid in the Germano coordinates. Following representation of Kühnen et al (2014, 2015), the critical parameters are plotted versus the radii ratio of the pipe and its centerline in Fig. 6. For the toroidal pipe the radii ratio coincide with the dimensionless curvature $\varepsilon$. Values of the dimensionless curvature and torsion corresponding to the experimental points of Kühnen et al (2015) are shown in Fig. 6a as a table.

For $\varepsilon \geq 0.02$, the critical Reynolds numbers for toroidal pipes obtained using all the three formulations coincide between themselves to within the second decimal place and agree well with the results of Canton et al (2016). Since no convergence study was presented in the latter work, we cannot estimate accuracy of the results reported there. At very small curvatures (radii ratio), $\varepsilon < 0.01$, the results of Canton et al (2016) show a steep decrease of $Re_{cr}$ with the decrease of the curvature from approximately 0.017 to 0.01, and then a steep increase of $Re_{cr}$ with further decrease of the radii ratio (Fig. 6). As noted by Canton et al (2016), this growth corresponds to the infinite $Re_{cr}$ in the straight pipe, which is the limiting case corresponding to $\varepsilon \to 0$. In this work we did not repeat the calculation of the whole critical curve of the toroidal pipe, and only verified several critical points starting from the neutral curve minimum at $\varepsilon = 0.01$.

For the above mentioned curvature $\varepsilon = 0.01$, the critical Reynolds numbers calculated by the formulations 1 and 2 were 4197 and 4184, respectively. Calculation in the cylindrical coordinates yielded 4238. The result of Catton et al (2016) was 4257. The critical azimuthal wavenumber was $m = 74$ ($k = 0.74$) in calculations of Catton and in the present calculations in the cylindrical coordinates, and $m = 73$ ($k = 0.73$) for calculation in the Germano coordinates. It should be noted that the critical Reynolds numbers for $m = 73$ and 74 differ only in the fourth decimal place, and can be considered as numerically indistinguishable. The critical frequencies calculated by the formulations 1 and 2 were -1.720, by formulation in the cylindrical coordinates -1.743 for $m = 74$ and -1.720 for $m = 73$, and the value of Catton et al (2016) is -1.736 for $m = 74$.

The agreement between the critical frequencies calculated for the toroidal pipe is as good as for the critical Reynolds numbers (Fig. 6b). Note that all the calculated frequencies are negative. In this study the most unstable perturbations in critical points are assumed to behave as $exp[i(\omega t + ks)]$. Thus, the negative $\omega$ means that the instability sets in as a travelling wave propagating upstream. Canton et al (2016) assumed that the perturbations behave as $exp[i(k\varphi - \omega t)]$ and reported positive critical frequencies, thus arriving to the same conclusion.

Comparing with the experimental results of Kühnen et al (2014, 2015), we observe that experimental critical Reynolds numbers are slightly larger than the computed ones. This can be expected since oscillations can be detected in an experiment only at some finite amplitude, while the numerical critical Reynolds number corresponds to the limit of zero amplitude. Note, that experiments in the helical pipe were carried out at different torsions, so that these experimental points do not belong to the same neutral curve. Nevertheless, at small torsions the critical Reynolds numbers of the helical pipe are quite close to those of the toroidal one, which is also the case for numerical results (Fig. 6a). This is not observed, however, in the graph of critical frequencies (Fig. 6b). For the helical and toroidal pipes they have comparable values, but exhibit qualitatively different dependencies on the radii ratio, which is common for the experimental and numerical results. At the same time the disagreement between the experimental and numerical values is noticeably larger than for the critical Reynolds numbers.

Comparison with earlier experiments that addressed onset of instability in helical pipes is not as good as the above recent studies of Kühnen et al. (2014, 2015). One of the earliest experiments on instability was conducted by White (1929) and Taylor (1929). Much later this instability was studied experimentally by Sreenivasan & Strykowski (1983), Webster & Humphrey (1993) and Yamamoto (1995), and numerically by Yamamoto et al (1998) and Di Piazza & Ciofalo (2011).

White (1929), Taylor (1929), Sreenivasan & Strykowski (1983) and Webster & Humphrey (1993) studied helical pipes with very small torsions with the torsion to curvature ratio $\lambda < 0.06$. Our computations for parameters of these experiments were very close to the $\lambda = 0$ results obtained for the toroidal pipe (Fig. 6a). For the pipes with a very small curvature, $\varepsilon = 0.00049$ in the experiments of White (1929), and $\varepsilon < 0.0072$ in Sreenivasan & Strykowski (1983), the calculated critical Reynolds numbers are noticeably larger than the observed ones, which correspond to the bypass transition similar to one occurring in the straight circular pipe. However, in the experiments with larger curvatures, $\varepsilon = 0.02$ and 0.066 in White (1929), $\varepsilon = 0.031$ and 0.06 in Taylor (1929), $0.02 < \varepsilon < 0.12$ in Sreenivasan & Strykowski (1983), $\varepsilon = 0.055$ in Webster & Humphrey (1993), the observed critical Reynolds numbers are above 5000 and are noticeably larger than those calculated in the present study and in Canton et al (2016) for the toroidal pipe. We cannot judge the accuracy of the above experiments. At the same time, we note that critical Reynolds numbers in these experiments were defined as location of the first break at the dependence of the friction factor on the Reynolds number (White, 1929), or just by visual observation of the flow (Taylor, 1929; Sreenivasan & Strykowski, 1983). This can be much less precise than the direct pointwise LDV measurements carried out by Kühnen et al. (2014, 2015). Webster & Humphrey (1993) also applied LDV measurements and for the curvature $\varepsilon \approx 0.05$ reported appearance of oscillations at $Re = 5060$ with the dimensionless frequency of ≈0.24. For a similar curvature and working liquid, Kühnen et al. (2015) reported oscillations starting form a noticeably smaller $Re = 3870$ with an order of magnitude larger value of the dimensionless frequency of ≈2.8, which reduces to a smaller value of ≈2.4 at $Re = 5040$, so that at close

Reynolds numbers, the frequencies measured in the two experiments differ in 10 times. Present calculations using the geometric parameters of the above experiments yielded the critical Reynolds numbers close to those obtained for the toroidal pipe, as well as those reported in the experiments of Kühnen et al. (2014, 2015). Apparently, small torsions do not affect noticeably the values of $Re_{cr}$.

Di Piazza & Ciofalo (2011) carried out fully non-linear three-dimensional time-dependent calculations for a toroidal pipe and found for the radii ratio $a/c = 0.1$ and 0.3, respectively, $Re_{cr} = 5175$ and 4575. The results of Canton et al. (2016) for these curvatures are, respectively, 3330 and 3024, while the present results are, respectively, 3376 and 3045. The most possible reason for such overestimation of the critical Reynolds numbers is under resolution of the most unstable perturbation in the three-dimensional computation.

Instabilities in helical pipes with a large torsion were studied experimentally in Yamamoto et al (1995) and numerically in Yamamoto et al (1998). The measurements were carried out by a hot wire. The numerical study considered only *s*-independent perturbations, and reported intervals of the Reynolds number inside which the instability onset is expected. In spite of incomplete mathematical model, the authors succeeded to establish a qualitative agreement between their calculations and measurements. To compare with these experiments, we picked up several points from Fig. 9 of Yamamoto et al (1995), denoted there as black triangles that indicated cases "where the signal showed a wavy but not irregular structure". The results are reported in Table 6. The column corresponding to the above mentioned triangles is denoted as Y1995. The neighbor column corresponding to the experimental intervals between observed steady and unsteady flow states, taken form Fig. 7 of Yamamoto et al (1998), is denoted as Y1998. We observe that the present critical Reynolds numbers are either inside the reported intervals, or below them. An interesting observation is that when the calculated $Re_{cr}$ fall inside the experimentally predicted interval, their values are very close to those indicated by the black triangles in Yamamoto et al (1995). These two cases are shown in bold (Table 6). In all the other cases, the calculated critical Reynolds numbers are below the experimental intervals. It should be noted that smaller calculated critical Reynolds numbers were reported also in Yamamoto et al (1998). Due to the present complete formulation, the present $Re_{cr}$ are even smaller than those reported in Yamamoto et al (1998).

### 5.5. *An example of critical curve*

To illustrate an outcome of a parametric stability study we present an example in Fig. 7, where we consider a helical pipe with the curvature $\varepsilon = 0.05$ and vary the torsion to curvature ratio $\lambda$ from zero to 5. At the neutral curve $Re_{cr}(\lambda)$, shown in Fig. 7a, we observe decrease of the critical Reynolds number with the increasing torsion, reported also in Yamamoto et al (1995, 1998) and above for a larger curvature. The inserts in Fig. 7a illustrate how the shape of the Dean vortices changes along the curve. The neutral curve, in fact, is the lower envelope of three different marginal curves belonging to three different eigenvectors of the linear stability problem. The latter is clearly seen in Fig. 7b where the critical circular frequency is shown together with the critical wavenumber as functions of $\lambda$. Each dependence $\omega_{cr}(\lambda)$ and $k_{cr}(\lambda)$ is represented by three separate curves belonging to the three eigenmodes. The intervals in which these curves are plotted correspond to the intervals of $\lambda$ where these modes are dominant (most unstable). Within the linear stability model, these modes replace each other abruptly, but they may grow together and interact in the fully non-linear regime.

Inserts in Fig. 7b show patterns of the absolute value of perturbation of the velocity $v_s$. As explained above, these patterns represent distribution of the oscillations amplitude in a slightly supercritical regime. At $\lambda = 0$, the Dean vortices of the base flow are antisymmetric, while the most unstable perturbation exhibits a symmetric pattern. Also, the real and imaginary parts of the perturbation (not shown here) are symmetric. Altogether, it means that the vortices oscillate in a counter phase, but with the same amplitudes. With the growth of $\lambda$, the upper counter clockwise rotating vortex becomes stronger and at $\lambda \approx 0.5$ another eigenmode producing significantly stronger oscillations of the upper vortex becomes most unstable. The characteristic oscillations amplitude is shown by an insert for $\lambda = 1$ (Fig. 7b). At $\lambda > 1$ the third eigenmode, which excites oscillations in the lower vortex, becomes most unstable. To show that the eigenmode exhibits similar amplitude patterns at different torsions, it is plotted for both $\lambda = 3$ and 4.

A complete stability study must include parametric studies over all relevant values of $\varepsilon$ and $\lambda$ with a more detail examination of physical reasons leading to instability. This would require a large amount of computational runs, as well as much additional graphics. This is beyond the scope of the present paper.

## 4. Concluding remarks

The primary goal of this study was to establish an accurate and robust computational model to study stability of steady flows in helical pipes. The computations were carried out using two different formulations with an additional third formulation applied for a toroidal pipe. As noted, the toroidal pipe is a limiting case of the helical one with vanishing torsion.

In the first part of this study we successfully compared steady states calculated by all three approaches with the experimental measurements of De Amicis et al (2014) and with the numerical results of Yamamoto et al. (1994). We also presented convergence studies and some additional data for possible further comparisons.

The second part of the results relates to the linear instability of steady flows. Since there is much less independent data for a quantitative comparison, we paid much attention to the convergence of eigenvalues and critical parameters. Then we carried out computations for parameters of recent experimental studies of Kühnen et al. (2014, 2015) and compared them with the published stability results for the toroidal pipe (Canton et al, 2016). The comparisons with the independent calculations were good, while comparisons with the experiment showed a good agreement in the values of the critical Reynolds numbers, but only qualitative agreement in the values of the critical frequencies.

Attempts to compare with earlier experimental stability studies (White, 1929; Taylor, 1929; Sreenivasan & Strykowski, 1983; Webster & Humphrey, 1993) showed that their results are noticeably larger than those obtained here or measured in Kühnen et al. (2014, 2015) for similar pipe curvatures. In all these experiments, the pipe torsions were too small to significantly affect the critical Reynolds numbers. Thus, the critical Reynolds numbers calculated for parameters of the above experiments appeared to be close to those reported in Kühnen et al. (2014, 2015).

The study of Yamamoto (1995) allowed us to compare also with the experimental results obtained for helical pipes with large torsion. Only in two out of eight cases, taken for the comparison, the computed critical Reynolds numbers laid inside the experimentally predicted intervals and agreed well with the experimental values corresponding to the appearance of regular oscillations. In the six other cases, the calculated critical values were below the experimentally predicted intervals.

To illustrate complicacy of a complete parametric stability study yet to be done, we reported an example of the stability results for a single fixed value of the pipe curvature, $\varepsilon = 0.05$, and torsion to curvature ratio varied from zero to 5. These results show that depending on both geometric parameters, curvature and torsion, the instability is caused by different most unstable perturbations, which leads also to qualitatively different oscillatory flow regimes. Therefore, a thorough parametric study that considers all relevant ranges of the curvature and the torsion is needed, but is beyond the scope of the present study. However, already on the basis of current results we can confirm that the critical Reynolds number steeply decreases with the increase of torsion, which we observed for small ($\varepsilon = 0.05$) and large ($\varepsilon = 0.4$) curvatures. We have found also that in most of the cases considered, the travelling wave developing due to the primary instability propagates upstream.

Returning to the experimentally measured critical Reynolds numbers that appear to be significantly larger than the calculated ones (see, e.g., Table 6), we point on two possible reasons. First, the instabilities studied here numerically are temporal. Instabilities in experimental setups, which necessarily have an entrance and an exit, develop spatially. The temporal instabilities obtained numerically set in as a wave traveling along the helical centerline. In most of the cases considered, and always at small torsions, the waves travel upstream. To make the experiment "close" to the temporal formulation, the pipe must be sufficiently long and the waves should not be suppressed at the entrance and the exit. This may not be the case in all the experiments. The second reason can be assumed from the distribution of oscillation amplitudes shown in the inserts of Fig. 7b. If pointwise measurements, e.g., LDV or hot wire are carried out in the region where the amplitude is small or even vanishing, no instability will be detected. Numerical results, like those presented here, can help in planning of such experiments as it was done for another problem in Haslavsky et al (2011).

Acknowledgments:

This research was supported by Israel Science Foundation (ISF) grant No 415/18 and was enabled in part by support provided by WestGrid ([www.westgrid.ca](www.westgrid.ca)) and Compute Canada ([www.computecanada.ca](www.computecanada.ca)). The author would like to express his thanks to A. Cammi, J. Canton and J. Kühnen for kindly agreeing to share their results.

# Figure captions

**Figure 1.** Sketch of a helical pipe (left) and illustration of helical coordinates introduced by Germano (1982).

**Figure 2.** Sketch of a cross-section of torus in the cylindrical coordinates $(\hat{r}, \varphi, z)$.

**Figure 3.** Comparison of measured and calculated friction factors. Lines – calculations, symbols – results of De Amicis et al. (2014).

**Figure 4.** Isolines of the $s$ −velocity and pseudo – streamfunction $\psi$ for parameters of Yamamoto et al. (1994), $Dn = 1000, \delta = 0.4$, and $0 \leq \delta \leq 0.6$. The inner side of the pipe is on the left.

**Figure5.** Isolines of the $s$ −velocity and pseudo – streamfunction $\psi$ for parameters of Yamamoto et al. (1994), $Dn = 1000, \delta = 0.4$, and $1 \leq \delta \leq 2.5$. The inner side of the pipe is on the left.

**Figure 6.** Comparison of present linear stability results with the experimental results of Kühnen et al (2014, 2015) and numerical results of Canton et al (2016).

**Figure 7.** Neutral curve $Re_{cr}(\lambda)$ for $\varepsilon = 0.05$ (a), and corresponding dependencies of $\omega_{cr}(\lambda)$ (black) and $k_{cr}(\lambda)$ (red). The inserts in frame (a) show isolines of the pseudo-streamfunction $\psi$. Inserts in frame (b) show amplitude of the most unstable perturbations of $v_s$.

**Appendix**

For the "two-dimensional" flow depending only on the coordinates $r$ and $\xi$, the momentum equation in an alternative form with the eliminated mixed second derivatives reads

$$\frac{\partial v_r}{\partial t} + v_r \frac{\partial v_r}{\partial r} + \frac{v_\xi}{r}\frac{\partial v_r}{\partial \xi} - \varepsilon\lambda \frac{v_s}{H_s}\frac{\partial v_r}{\partial \xi} - \frac{v_\xi^2}{r} - \frac{\varepsilon \sin(\xi)}{H_s}v_s^2 =$$

$$= -\frac{\partial p}{\partial r} + \frac{1}{R_G}\frac{1}{rH_s}\left\{\frac{\partial^2}{\partial r^2}[rH_s v_r] + \frac{1}{r}\frac{\partial}{\partial \xi}\left[H_s \frac{\partial v_r}{\partial \xi}\right] + \varepsilon^2\lambda^2 r \frac{\partial}{\partial \xi}\left[\frac{1}{H_s}\frac{\partial v_r}{\partial \xi}\right] - \frac{1}{r}\frac{\partial v_\xi}{\partial \xi} - \varepsilon\lambda \frac{\partial}{\partial \xi}\left[\frac{v_s}{H_s}\right]\right\} \quad (A1)$$

$$\frac{\partial v_\xi}{\partial t} + v_r \frac{\partial v_\xi}{\partial r} + \frac{v_\xi}{r}\frac{\partial v_\xi}{\partial \xi} - \varepsilon\lambda \frac{v_s}{H_s}\frac{\partial v_\xi}{\partial \xi} + \frac{v_\xi v_s}{r} - \frac{\varepsilon \cos(\xi)}{H_s}v_s^2 = -\frac{1}{r}\frac{\partial p}{\partial \xi} + \frac{1}{R_G}\frac{1}{H_s}\left\{\frac{\partial}{\partial r}\left[\frac{H_s}{r}\frac{\partial(rv_\xi)}{\partial r}\right] +\right.$$

$$\left.+ \frac{1}{r^2}\frac{\partial^2}{\partial \xi^2}[H_s v_\xi] + \varepsilon^2\lambda^2 \frac{\partial}{\partial \xi}\left[\frac{1}{H_s}\frac{\partial v_\xi}{\partial \xi}\right] + \frac{2}{r^2}\frac{\partial(H_s v_r)}{\partial \xi} + \varepsilon \cos(\xi)\frac{\partial v_r}{\partial r} + \varepsilon^2\lambda \frac{\partial}{\partial \xi}\left[\frac{\cos(\xi)}{H_s}v_s\right]\right\} \quad (A2)$$

$$\frac{\partial v_s}{\partial t} + v_r \frac{\partial v_s}{\partial r} + \frac{v_\xi}{r}\frac{\partial v_s}{\partial \xi} - \varepsilon\lambda \frac{v_s}{H_s}\frac{\partial v_s}{\partial \xi} + \frac{\varepsilon \sin(\xi)}{H_s}v_r v_s + \frac{\varepsilon \cos(\xi)}{H_s}v_\xi v_s = -\frac{1}{H_s}\frac{\partial p}{\partial s} +$$

$$\frac{1}{R_G}\left\{\frac{1}{r}\frac{\partial}{\partial r}\left[\frac{r}{H_s}\frac{\partial(H_s v_s)}{\partial r}\right] + \frac{1}{r^2}\frac{\partial}{\partial \xi}\left[\frac{1}{H_s}\frac{\partial(H_s v_s)}{\partial \xi}\right] + \varepsilon^2\lambda^2 \frac{\partial}{\partial \xi}\left[\frac{1}{H_s^2}\frac{\partial v_s}{\partial \xi}\right] + \frac{\varepsilon^2 \lambda \cos(\xi)}{r}\frac{\partial}{\partial r}\left[\frac{r^2}{H_s^2}v_r\right] -$$

$$2\varepsilon^2\lambda \frac{\partial}{\partial \xi}\left[\frac{\sin(\xi)}{H_s^2}v_r\right] - \varepsilon^2\lambda \frac{\partial}{\partial \xi}\left[\frac{\cos(\xi)}{H_s}v_\xi\right]\right\} \quad (A3)$$

The equations linearized near the steady state flow $\{V_r(r,\xi), V_\xi(r,\xi), V_s(r,\xi), P(r,\xi)\}$ that govern infinitely small disturbances $\{u_r(r,\xi), u_\xi(r,\xi), u_s(r,\xi), p(r,\xi)\}exp[i\sigma t + iks]$ are

$$\lambda u_r + V_r \frac{\partial u_r}{\partial r} + u_r \frac{\partial V_r}{\partial r} + \frac{V_\xi}{r}\frac{\partial u_r}{\partial \xi} + \frac{u_\xi}{r}\frac{\partial V_r}{\partial \xi} - \varepsilon\lambda u_s \frac{\partial V_r}{\partial \xi} + \frac{V_s}{H_s}\left[-\varepsilon\lambda \frac{\partial u_r}{\partial \xi} + iku_r\right] - \frac{2V_\xi u_\xi}{r} -$$

$$\frac{2\varepsilon \sin(\xi)}{H_s}V_s u_s = -\frac{\partial p}{\partial r} + \frac{1}{R_G}\frac{1}{rH_s}\left\{\frac{\partial^2}{\partial r^2}[rH_s u_r] + \frac{1}{r}\frac{\partial}{\partial \xi}\left[H_s \frac{\partial u_r}{\partial \xi}\right] + \varepsilon^2\lambda^2 r \frac{\partial}{\partial \xi}\left[\frac{1}{H_s}\frac{\partial u_r}{\partial \xi}\right] - \frac{1}{r}\frac{\partial u_\xi}{\partial \xi} -\right.$$

$$\left.\varepsilon\lambda \frac{\partial}{\partial \xi}\left[\frac{u_s}{H_s}\right] - k^2 \frac{ru_r}{H_s} - ik\varepsilon\lambda r\frac{\partial}{\partial \xi}\left[\frac{u_r}{H_s}\right] - \frac{ikr}{H_s}\varepsilon\lambda \frac{\partial u_r}{\partial \xi} + \frac{ik}{H_s}u_s\right\} \quad (A4)$$

$$\lambda u_\xi + V_r \frac{\partial u_\xi}{\partial r} + u_r \frac{\partial V_\xi}{\partial r} + \frac{V_\xi}{r}\frac{\partial u_\xi}{\partial \xi} + \frac{u_\xi}{r}\frac{\partial V_\xi}{\partial \xi} - \varepsilon\lambda \frac{V_s}{H_s}\frac{\partial u_\xi}{\partial \xi} - \varepsilon\lambda \frac{u_s}{H_s}\frac{\partial V_\xi}{\partial \xi} + \frac{V_\xi u_s}{r} + \frac{u_\xi V_s}{r} -$$

$$\frac{2\varepsilon \cos(\xi)}{H_s}V_s u_s + \frac{ik}{H_s}V_s u_\xi = -\frac{1}{r}\frac{\partial p}{\partial \xi} + \frac{1}{R_G}\frac{1}{H_s}\left\{\frac{\partial}{\partial r}\left[\frac{H_s}{r}\frac{\partial(ru_\xi)}{\partial r}\right] + + \frac{1}{r^2}\frac{\partial^2}{\partial \xi^2}[H_s u_\xi] + \varepsilon^2\lambda^2 \frac{\partial}{\partial \xi}\left[\frac{1}{H_s}\frac{\partial u_\xi}{\partial \xi}\right] +\right.$$

$$\left.\frac{2}{r^2}\frac{\partial(H_s u_r)}{\partial \xi} + \varepsilon \cos(\xi)\frac{\partial u_r}{\partial r} + \varepsilon^2\lambda \frac{\partial}{\partial \xi}\left[\frac{\cos(\xi)}{H_s}u_s\right] - k^2 \frac{u_\xi}{H_s} - ik\varepsilon\lambda \frac{\partial}{\partial \xi}\left[\frac{u_\xi}{H_s}\right] - \frac{ik}{H_s}\varepsilon\lambda \frac{\partial u_\xi}{\partial \xi} - ik\varepsilon \cos(\xi)\frac{u_s}{H_s}\right\}$$

$$(A5)$$

$$\lambda v_s + V_r \frac{\partial u_s}{\partial r} + u_r \frac{\partial V_s}{\partial r} + \frac{V_\xi}{r}\frac{\partial u_s}{\partial \xi} + \frac{u_\xi}{r}\frac{\partial V_s}{\partial \xi} - \varepsilon\lambda \frac{V_s}{H_s}\frac{\partial u_s}{\partial \xi} - \varepsilon\lambda \frac{u_s}{H_s}\frac{\partial V_s}{\partial \xi} + \frac{\varepsilon \sin(\xi)}{H_s}(V_r u_s + u_r V_s) +$$
$$\frac{\varepsilon \cos(\xi)}{H_s}\left(V_\xi u_s + u_\xi V_s\right) + \frac{ik}{H_s} V_s u_s = \frac{\varepsilon\lambda}{H_s}\frac{\partial p}{\partial \xi} - \frac{ik}{H_s} p + \frac{1}{R_G}\left\{\frac{1}{r}\frac{\partial}{\partial r}\left[\frac{r}{H_s}\frac{\partial (H_s u_s)}{\partial r}\right] + \frac{1}{r^2}\frac{\partial}{\partial \xi}\left[\frac{1}{H_s}\frac{\partial (H_s u_s)}{\partial \xi}\right] +\right.$$
$$\varepsilon^2\lambda^2 \frac{\partial}{\partial \xi}\left[\frac{1}{H_s^2}\frac{\partial u_s}{\partial \xi}\right] + \frac{\varepsilon^2 \lambda \cos(\xi)}{r}\frac{\partial}{\partial r}\left[\frac{r^2}{H_s^2} u_r\right] - 2\varepsilon^2\lambda \frac{\partial}{\partial \xi}\left[\frac{\sin(\xi)}{H_s^2} u_r\right] - \varepsilon^2\lambda \frac{\partial}{\partial \xi}\left[\frac{\cos(\xi)}{H_s} u_\xi\right] - \frac{ik}{r}\frac{\partial}{\partial \xi}\left[\frac{r}{H_s} u_\xi\right] -$$
$$\left.\frac{ik}{r}\frac{\partial}{\partial r}\left[\frac{r}{H_s} u_r\right]\right\} \quad (A6)$$

Table 1. Values of $Q/Q_{Poiseuille}$ for the parameters $Dn=1000$, $\beta=0.4$ of Yamamoto et al (1994). Convergence using the two formulations and comparison with the results of Yamamoto et al (1994). Uniform grid.

| δ | Yamamoto et al. (1994) | Formulation 1 | | | Formulation 2 | | |
|---|---|---|---|---|---|---|---|
| | | 50×100 | 100×200 | **Richardson extrapolation** | 50×100 | 100×200 | **Richardson extrapolation** |
| 0 | 0.5995 | 0.5993 | 0.5994 | 0.5995 | 0.5993 | 0.5994 | 0.5995 |
| 0.2 | 0.5917 | 0.5915 | 0.5916 | 0.5917 | 0.5915 | 0.5917 | 0.5919 |
| 0.4 | 0.5705 | 0.5704 | 0.5705 | 0.5706 | 0.5704 | 0.5704 | 0.5704 |
| 0.6 | 0.5424 | 0.5425 | 0.5425 | 0.5425 | 0.5425 | 0.5425 | 0.5425 |
| 0.8 | 0.5208 | 0.5226 | 0.5227 | 0.5228 | 0.5226 | 0.5226 | 0.5226 |
| 1.0 | 0.5600 | 0.5719 | 0.5722 | 0.5725 | 0.5719 | 0.5721 | 0.5723 |
| 1.2 | 0.6555 | 0.6742 | 0.6745 | 0.6748 | 0.67398 | 0.6743 | 0.6746 |
| 1.4 | 0.7312 | 0.7576 | 0.7578 | 0.7581 | 0.7572 | 0.7575 | 0.7578 |
| 1.6 | | 0.8211 | 0.8213 | 0.8215 | 0.82065 | 0.82098 | 0.8213 |
| 1.8 | | 0.8682 | 0.8683 | 0.86845 | 0.8677 | 0.86801 | 0.8683 |
| 2.0 | | 0.9027 | 0.9028 | 0.9029 | 0.9023 | 0.9026 | 0.9029 |
| 2.5 | | 0.9543 | 0.9543 | 0.9543 | 0.9538 | 0.9540 | 0.9543 |

Table 2. Minimal and maximal values of the stream function, and the friction factor, obtained as Richardson extrapolations using the uniform grids 50×100 and 100×200. The parameters $Dn=1000$, $\beta=0.4$ of Yamamoto et al (1994).

| $\delta$ | Formulation 1 | | | Formulation 2 | | |
|---|---|---|---|---|---|---|
| | $\psi_{min}$ | $\psi_{max}$ | $f$ | $\psi_{min}$ | $\psi_{max}$ | $f$ |
| 0   | -0.2435 | 0.2435   | 0.6372 | -0.2435 | 0.2435    | 0.6373 |
| 0.2 | -0.2599 | 0.2328   | 0.6542 | -0.2599 | 0.2325    | 0.6542 |
| 0.4 | -0.2847 | 0.2257   | 0.7035 | -0.2847 | 0.2257    | 0.7036 |
| 0.6 | -0.3261 | 0.2209   | 0.7778 | -0.3261 | 0.2209    | 0.7779 |
| 0.8 | -0.4137 | 0.2137   | 0.8380 | -0.4136 | 0.2137    | 0.8382 |
| 1.0 | -0.7261 | 0.1327   | 0.6986 | -0.7250 | 0.1326    | 0.6989 |
| 1.2 | -1.3149 | 0.03197  | 0.5029 | -1.3130 | 0.03180   | 0.5031 |
| 1.4 | -1.9316 | 0.006872 | 0.3978 | -1.9299 | 0.006865  | 0.3986 |
| 1.6 | -2.5122 | 0.000494 | 0.3402 | -2.5108 | 0.0004937 | 0.3394 |
| 1.8 | -3.0518 | 0        | 0.3036 | -3.0505 | 0         | 0.3037 |
| 2.0 | -3.5556 | 0        | 0.2808 | -3.5547 | 0         | 0.2810 |
| 2.5 | -4.7039 | 0        | 0.2514 | -4.7035 | 0         | 0.2515 |

Table 3. Results for the parameters $Dn=1000$, $\beta=0.4$ of Yamamoto (1994) computed for flow in a torus. To be compared with the results in Tables 1, 2 and 4 for $\delta = 0$.

| Grid | $Q/Q_{Poiseuille}$ | $|\psi|_{max}$ | $f$ | $\lambda(m=0)$ | $\lambda(m=2)$ |
|---|---|---|---|---|---|
| $100^2$ | 0.5590 | 0.2428 | 0.6381 | (-0.4887, 1.4657) | (-0.4508, -3.5894) |
| $200^2$ | 0.5994 | 0.2433 | 0.6374 | (-0.4884, 1.4667) | (-0.4505, -3.5916) |
| $300^2$ | 0.5995 | 0.2433 | 0.6368 | (-0.4885, 1.4669) | (-0.4508, -3.5923) |
| $400^2$ | 0.5995 | 0.2434 | 0.6371 | (-0.4886, 1.4668) | (-0.4512, -3.5925) |
| $500^2$ | 0.5995 | 0.2434 | 0.6371 | (-0.4887, 1.4668) | (-0.4513, -3.5924) |
| $600^2$ | 0.5995 | 0.2434 | 0.6371 | (-0.4887, 1.4668) | (-0.4513, -3.5925) |

Table 4. Leading eigenvalues calculated on the uniform grids 50×100, 100×200, 200×400 and 300×600.  The parameters $Dn$=1000, $\beta$=0.4 of [6].

| | Formulation 1, $\sigma(k=0)$ | | | | Formulation 2, $\sigma(k=0)$ | | | |
|---|---|---|---|---|---|---|---|---|
| $\delta$ | 50×100 | 100×200 | 200×400 | 300×600 | 50×100 | 100×200 | 200×400 | 300×600 |
| 0 | (-0.4891, 1.4652) | (-0.4889, 1.4658) | (-0.4888, 1.4659) | (-0.4888, 1.4659) | (-0.4890, 1.4655) | (-0.4888, 1.4657) | (-0.4887, 1.4658) | (-0.4887, 1.4658) |
| 0.5 | (-0.5447, 0) | (-0.5439, 0) | (-0.5443, 0) | (-0.5443,0) | (-0.5449, 0) | (-0.5448, 0) | (-0.5442, 0) | (-0.5442, 0) |
| 1.0 | (-0.1272, 2.7386) | (-0.1240, 2.7430) | (-0.1231, 2.7442) | (-0.1229, 2.7445) | (-0.1214, 2.7371) | (-0.1198, 2.7417) | (-0.1194, 2.7429) | (-0.1194, 2.7431) |
| 1.5 | (-0.2279, 0) | (-0.2278, 0) | (-0.2278, 0) | (-0.2277, 0) | (-0.2280, 0) | (-0.2279, 0) | (-0.2279, 0) | (-0.2279, 0) |
| 2.0 | (-0.1955, 0) | (-0.1954, 0) | (-0.1954, 0) | (-0.1954, 0) | (-0.1956, 0) | (-0.1955, 0) | (-0.1955, 0) | (-0.1954, 0) |
| 2.5 | (-0.1837, 0) | (-0.1837, 0) | (-0.1837, 0) | (-0.1837, 0) | (-0.1838, 0) | (-0.1837, 0) | (-0.1837, 0) | (-0.1837, 0) |

| | Formulation 1, $\sigma(k=0.8)$ | | | | Formulation 2, $\sigma(k=0.8)$ | | | |
|---|---|---|---|---|---|---|---|---|
| $\delta$ | 50×100 | 100×200 | 200×400 | 300×600 | 50×100 | 100×200 | 200×400 | 300×600 |
| 0 | (-0.4544, -3.5888) | (-0.4531, -3.5906) | (-0.4527, -3.5911) | (-0.4526, -3.5912) | (-0.4546, -3.5891) | (-0.4532, -3.5905) | (-0.4528, -3.5909) | (-0.4527, -3.5910) |
| 0.5 | (-0.5884, -0.8267) | (-0.5888, -0.8267) | (-0.5889, -0.8267) | (-0.5889, -0.8267) | (-0.5884, -0.8258) | (-0.5889, -0.8263) | (-0.5886, -0.8268) | (-0.5891, -0.8264) |
| 1.0 | (-0.2901, -6.7701) | (-0.2890, -6.7768) | (-0.2886, -6.7791) | (-0.2885, -6.7791)) | (-0.2846, -6.7685) | (-0.2849, -6.7755) | (-0.2851, -6.7773) | (-0.2851, -6.7777)) |
| 1.5 | (-0.005956, 2.1794) | (-0.006812, 2.1867) | (-0.006990, 2.1885) | (-0.007000, 2.1889) | (-0.01263, 2.1784) | (-0.01091, 2.1864) | (-0.01036, 2.1885) | (-0.01019, 2.1889) |
| 2.0 | ( -1.0028, 5.4529) | ( -1.0017, 5.4634) | ( -1.0016, 5.4663) | ( -1.0017, 5.4668) | ( -0.9891, 5.4502) | ( -0.9927, 5.4651) | ( -0.9939, 5.4690) | ( -0.9943, 5.4697) |
| 2.5 | (-1.0880, -5.3687) | (-1.0870, -5.3700) | (-1.0868, -5.3704) | (-1.0868, -5.3705) | (-1.0874, -5.3642) | (-1.0864, -5.3668) | (-1.0861, -5.3676) | (-1.0868, -5.3678) |

| | Torus, $\sigma(m=0)$ | | | | Torus, $\sigma(m=2)$ | | | |
|---|---|---|---|---|---|---|---|---|
| $\delta$ | 100×100 | 200×200 | 300×300 | 400×400 | 100×100 | 200×200 | 300×300 | 400×400 |
| 0 | (-0.4902, -1.4658) | (-0.4889, -1.4668) | (-0.4882, -1.4665) | (-0.4885, -1.4666) | (-0.45225, -3.5893) | (-0.4508, -3.5918) | (-0.4501 , -3.5922) | (-0.4506, -3.5923) |

Table 5. Critical parameters obtained as Richardson extrapolations using the uniform grids 300×600 and 400×800. $\varepsilon = 0.4$.

| $\beta$ | Formulation 1, 300×600 & 400×800 | | | | Formulation 2, 300×600 & 400×800 | | | |
|---|---|---|---|---|---|---|---|---|
| | $R_{G,cr}$ | $\omega_{cr}$ | $k_{cr}$ | $Re_{d,cr}$ | $R_{G,cr}$ | $\omega_{cr}$ | $k_{cr}$ | $Re_{d,cr}$ |
| 0 | 266.9632 | -8.9273 | 2 | 3279.37 | 266.982 | -8.9273 | 2 | 3279.72 |
| 0.5 | 214.0194 | -10.1205 | 1.050452 | 2223.81 | 213.889 | -10.10697 | 1.0489985 | 2221.94 |
| 1.0 | 55.1693 | -3.7646 | 0 | 353.167 | 54.7389 | -3.7575982 | 0 | 348.7052 |
| 1.5 | 32.4613 | 1.5235 | 0.9309 | 209.919 | 32.3842 | 1.52085 | 0.93053 | 208.962 |
| 2.0 | 59.4186 | -0.5673 | 5.3444 | 748.989 | 59.0670 | -0.58283 | 5.3444 | 739.862 |
| 2.5 | 52.082998 | 0.6937 | 2.1431 | 631.042 | 52.0163 | 0.70796 | 2.140764 | 629.372 |

| | Flow in torus by immersed boundary in cylindrical coordinates, $m=5$ | | | | | | | |
|---|---|---|---|---|---|---|---|---|
| | 100×100 | 200×200 | 300×300 | 400×400 | 500×500 | 600×600 | 700×700 | 800×800 |
| $R_{G,cr}$ | 254.706 | 264.123 | 264.755 | 265.563 | 265.689 | 265.949 | 266.219 | 266.441 |
| $\omega_{cr}$ | -9.0653 | -9.1187 | -9.0628 | -9.0265 | -9.0027 | -8.9926 | -8.9890 | -8.9733 |
| $k_{cr}$ | 2 | 2 | 2 | 2 | 2 | 2 | 2 | 2 |
| $Re_{d,cr}$ | 2994.12 | 3210.88 | 3232.73 | 3249.10 | 3251.64 | 3256.17 | 3258.70 | 3262.57 |

Table 6. Comparison with the experimental results of Yamamoto et el (1995).

| $\varepsilon$ | $\delta$ | $\lambda$ | $Re_{cr}$ from Y1995 | $Re_{cr}$ from Y1998 | $Re_{cr}$ present | $\omega_{cr}$ | $k_{cr}$ |
|---|---|---|---|---|---|---|---|
| **0.01** | **0.77** | **10.89** | **1680** | **1220 – 2050** | **1687** | **-10.31** | **0.9778** |
| 0.01 | 1.72 | 24.32 | 2280 | 1940 – 2440 | 1471 | 0.3717 | 0.6932 |
| 0.05 | 0.5 | 3.04 | 2560 | 2240 – 2860 | 1113 | -3.825 | 1.2884 |
| 0.05 | 0.92 | 5.82 | 1100 | 830 – 1380 | 626 | -1.866 | 0.8111 |
| 0.05 | 1.6 | 10.12 | 1100 | 900 – 1300 | 541 | 1.326 | 0.2656 |
| 0.1 | 0.45 | 2.01 | 2540 | 2280 – 2840 | 1573 | -2.9649 | 1.6048 |
| 0.1 | 0.74 | 3.31 | 1160 | 900 – 1670 | 371 | -8.326 | 2.1344 |
| **0.1** | **1.11** | **4.96** | **560** | **460 – 640** | **596** | **0.2517** | **0.4292** |

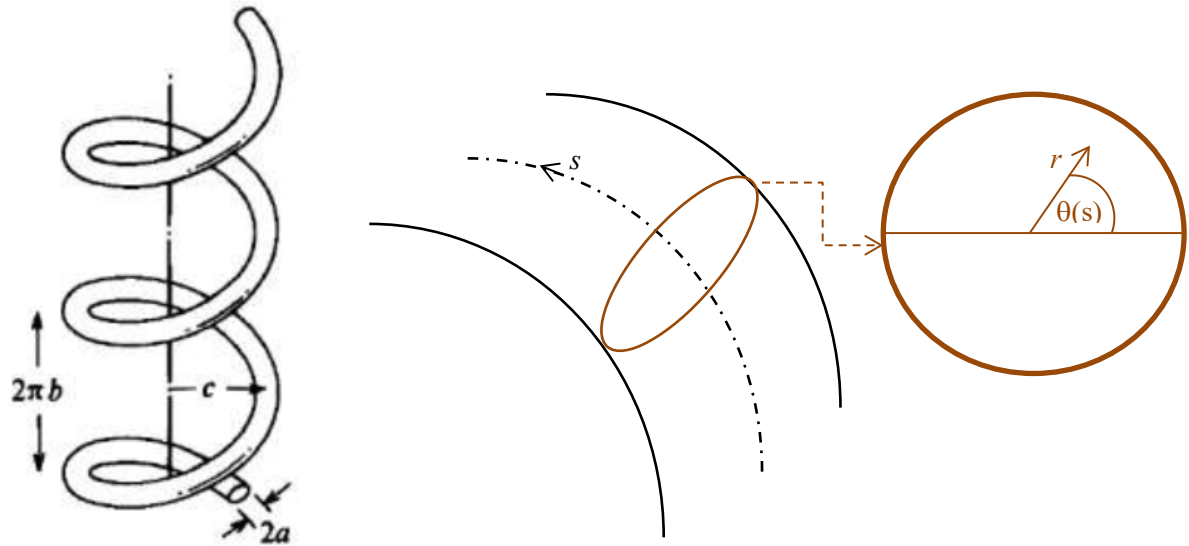

**Figure 1.** Sketch of a helical pipe (left, taken from Yamamoto, 1994) and illustration of helical coordinates introduced by Germano (1982).

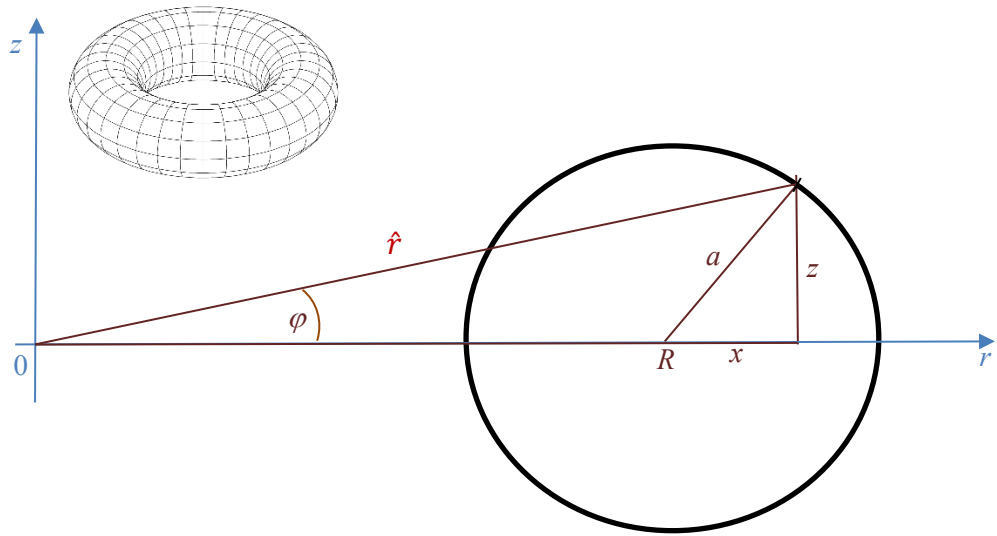

**Figure 2.** Sketch of a cross-section of torus in the cylindrical coordinates $(\hat{r}, \varphi, z)$.

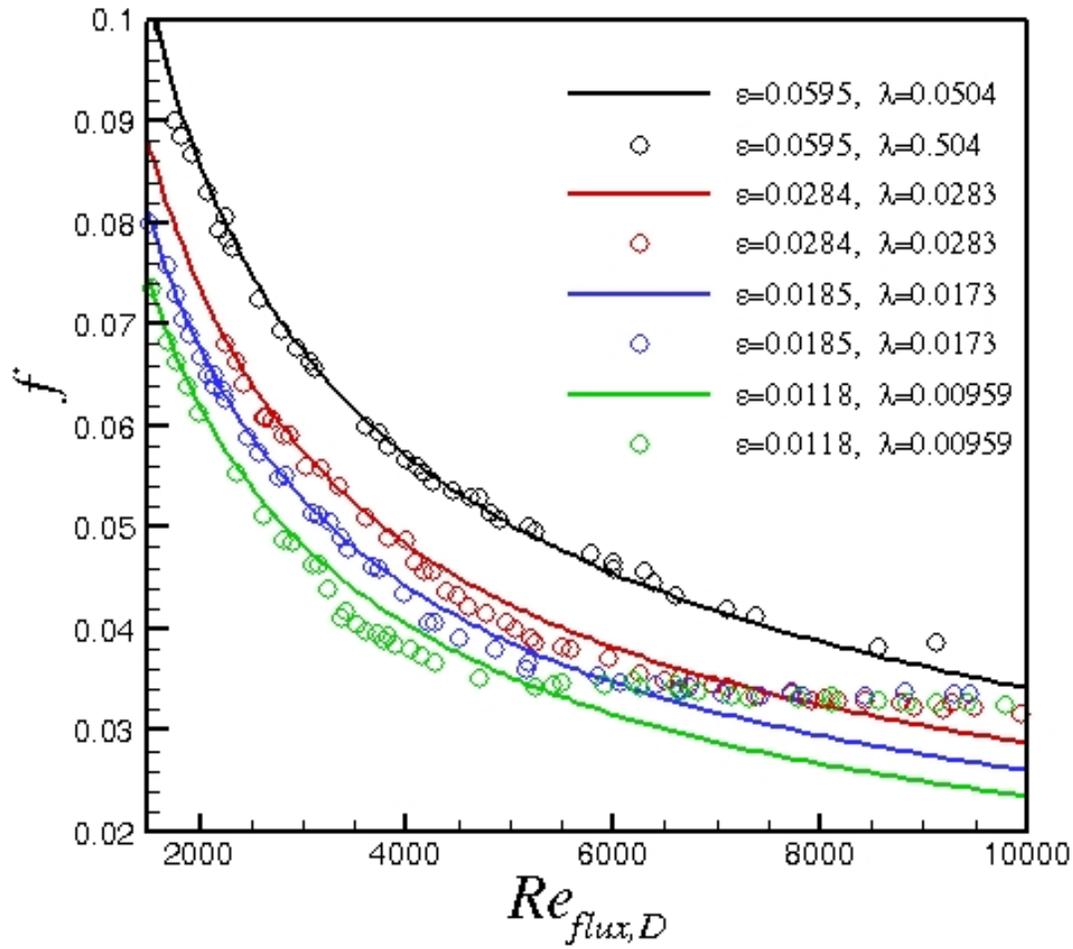

**Figure 3.** Comparison of measured and calculated friction factors. Lines – calculations, symbols – results of De Amicis et al. (2014)

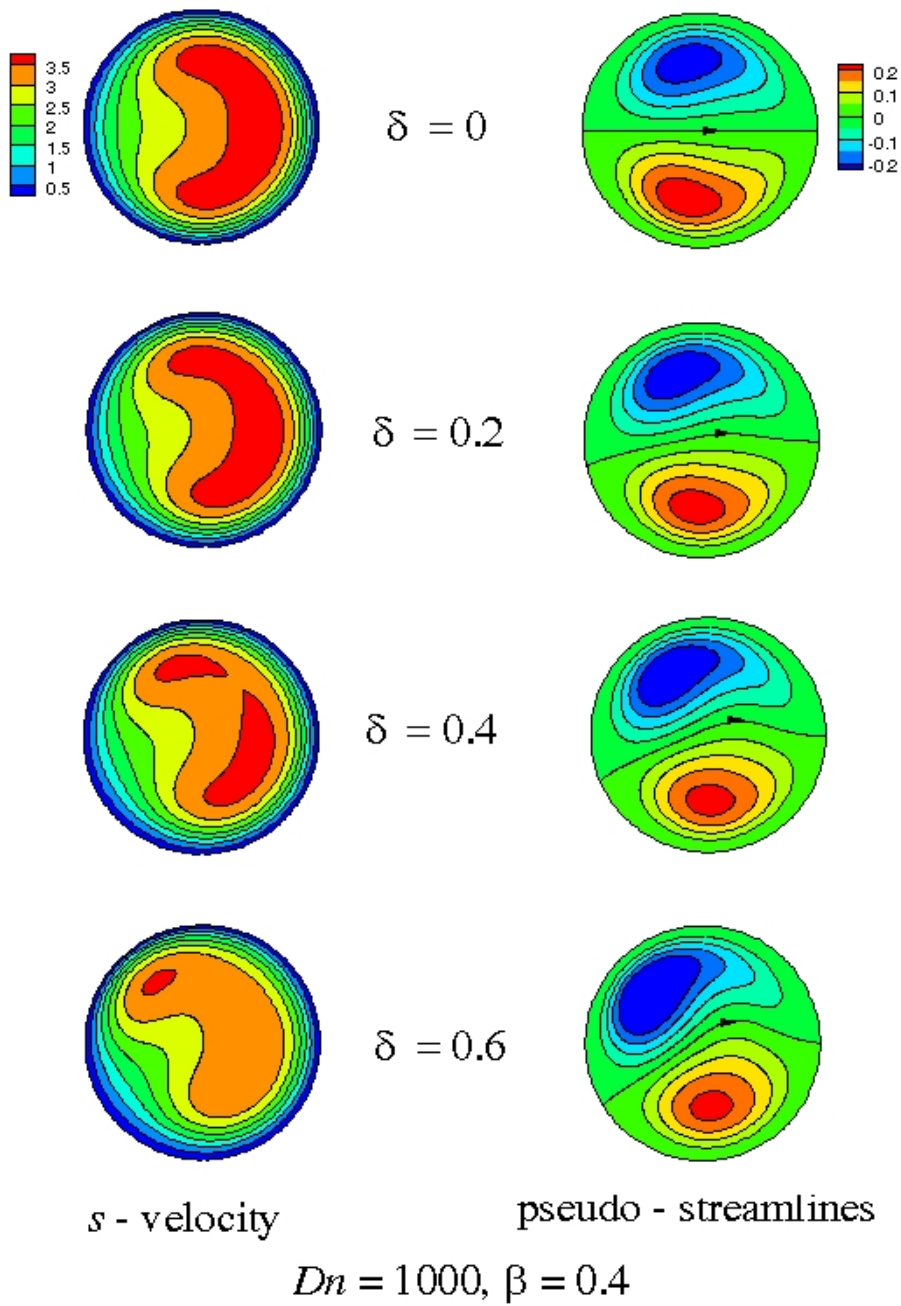

**Figure 4.** Isolines of the $s$ −velocity and pseudo – streamfunction $\psi$ for parameters of Yamamoto (1994), $Dn = 1000, \delta = 0.4$, and $0 \leq \delta \leq 0.6$. The inner side of the pipe is on the left.

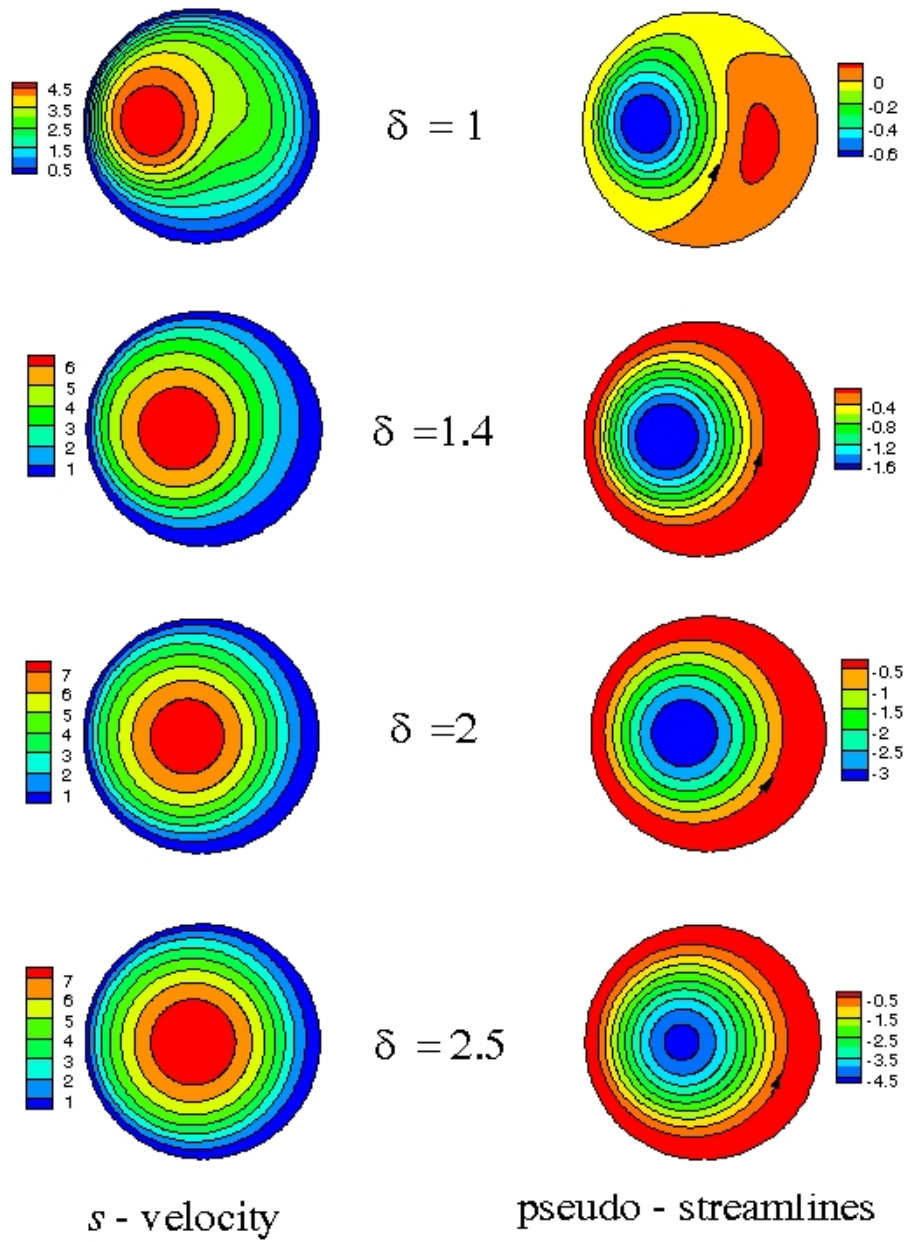

**Figure 5.** Isolines of the $s-$velocity and pseudo – streamfunction $\psi$ for parameters of Yamamoto (1994), $Dn = 1000, \delta = 0.4$, and $1 \leq \delta \leq 2.5$. The inner side of the pipe is on the left.

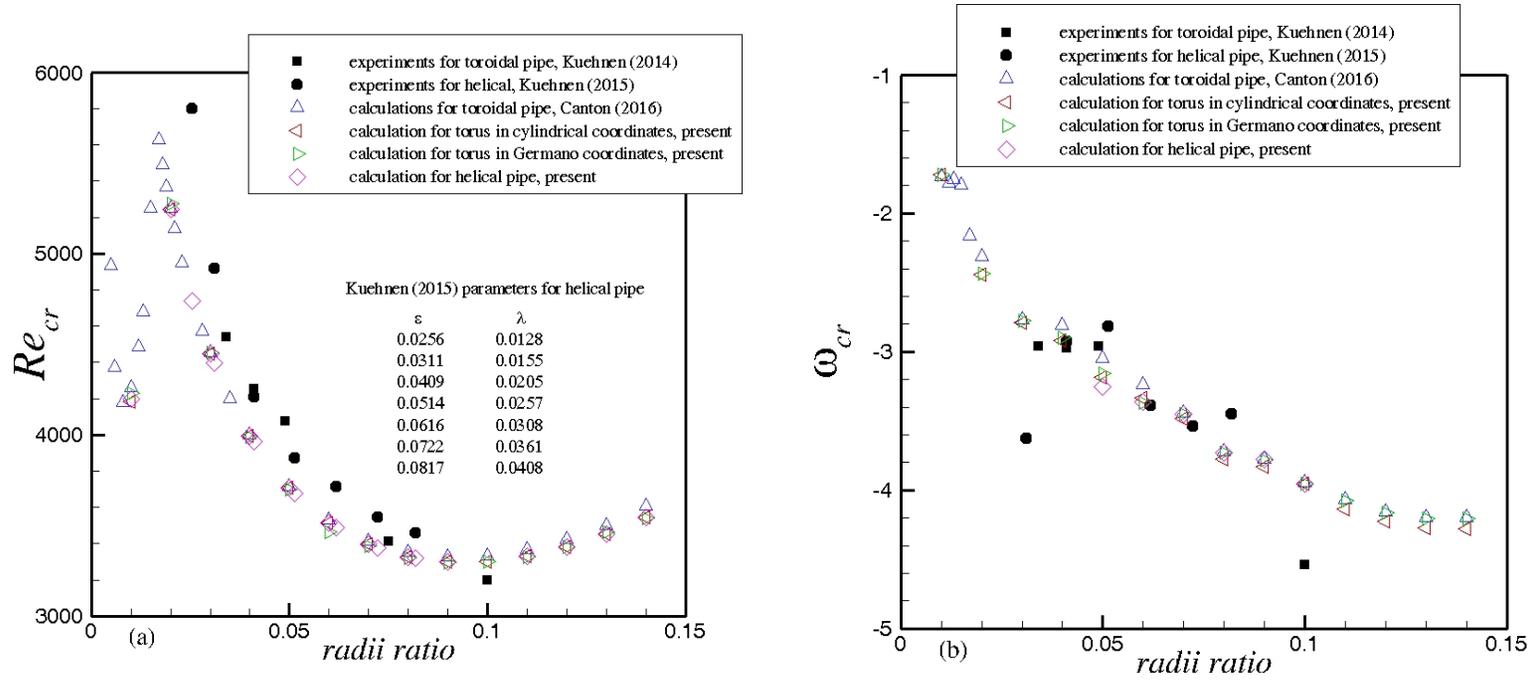

**Figure 6.** Comparison of present linear stability results with the experimental results of Kühnen et al (2014, 2015) and numerical results of Canton et al (2016).

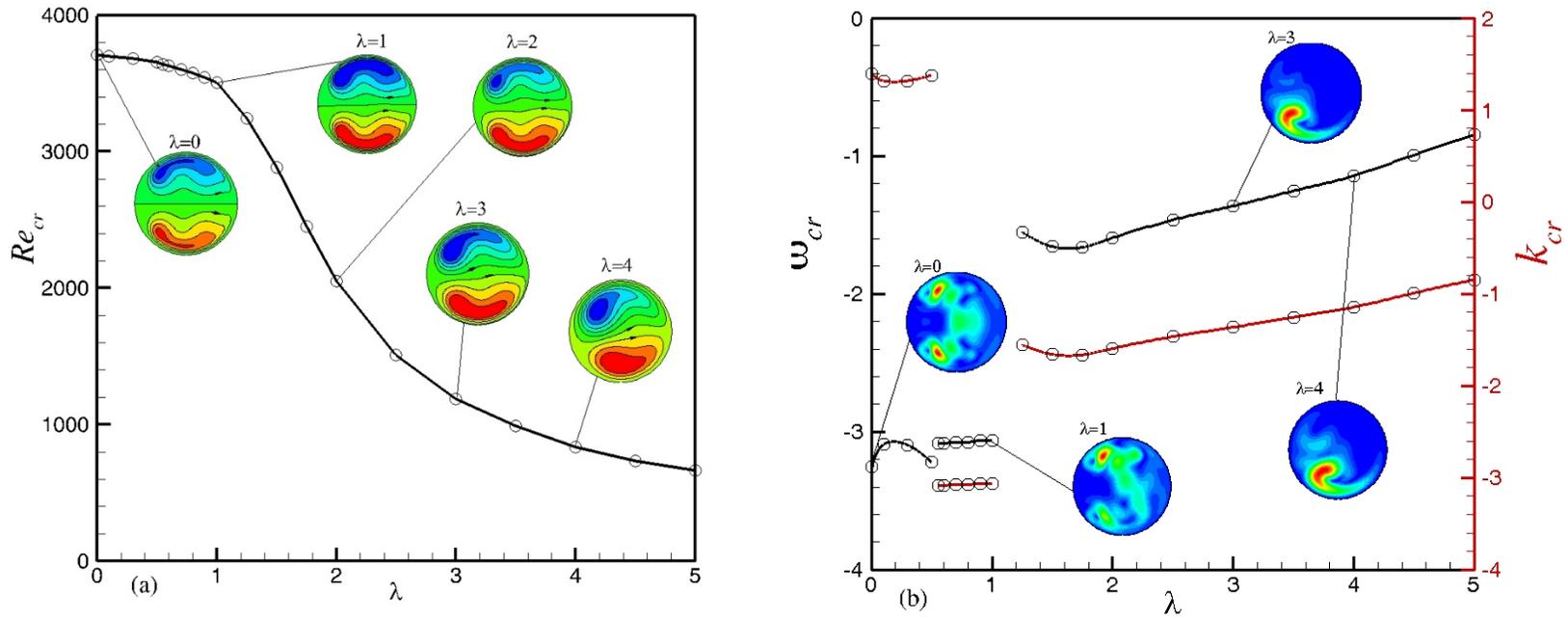

**Figure 7.** Neutral curve $Re_{cr}(\lambda)$ for $\varepsilon = 0.05$ (a), and corresponding dependencies of $\omega_{cr}(\lambda)$ (black) and $k_{cr}(\lambda)$ (red). The inserts in frame (a) show isolines of the pseudo-streamfunction $\psi$. Inserts in frame (b) show amplitude of the most unstable perturbations of $v_s$.